**SteatoNet- Modelling steatosis identifies candidate systemic flux distribution deregulations**

**Adviti Naik[1,2], Damjana Rozman[3], Aleš Belič[2*]**

[1] Faculty of Computer Sciences and Informatics, University of Ljubljana, Tržaška Cesta 25, 1000 Ljubljana, Slovenia

[2] Faculty of Electrical Engineering, University of Ljubljana, Tržaška Cesta 25, 1000 Ljubljana, Slovenia

[3] Centre for Functional Genomics and Bio-Chips, Institute of Biochemistry, Faculty of Medicine, University of Ljubljana, Ljubljana, Slovenia

*Corresponding author

Running title: SteatoNet: Modelling NAFLD-related steatosis

Manuscript length (without Methods): 43,386 characters

## ABSTRACT

The functional diversity in factors identified in association with non-alcoholic fatty liver disease (NAFLD) necessitates the utilization of holistic approaches to investigate the 'network' properties of NAFLD pathogenesis. We describe the generation of the first mouse-specific multi-tissue metabolic model, the SteatoNet, to investigate NAFLD-related steatosis, by utilising an object-oriented modelling approach and incorporating numerous hepatic metabolic pathways, their interaction with peripheral tissues and hierarchical feedback regulation at the transcriptional and post-translational level. The unique validated model is based on steady-state analysis of ordinary differential equations that allows investigation of systemic behaviour in the absence of estimated kinetic parameters. Model analysis indicated the importance of the metabolic flux distribution parameter in controlling metabolite concentration and identified critical focal points in the network that may play a pivotal role in initiating NAFLD. Albeit requiring experimental validation, the candidate triggers identified by SteatoNet provide insight into the network properties of NAFLD and emphasize the promising scope and potential of SteatoNet as a primary investigative tool for hypotheses generation and exploring related metabolic diseases.

## 1. INTRODUCTION

NAFLD, the hepatic manifestation of the metabolic syndrome, is the most common chronic liver disease among western populations, with a prevalence of 25-30% [1]. It encompasses a broad disease spectrum ranging from steatosis (hepatic accumulation of triglycerides), non-alcoholic steatohepatitis (NASH) that is additionally characterised by inflammation and progression into cirrhosis and hepatocellular carcinoma (HCC). Genome wide association studies (GWAS) studies have identified only a small number of genetic variants in association with NAFLD [2], which can be categorised into numerous different types of biological pathways such as inflammation, lipid and glucose metabolism etc. NAFLD has also been associated with aberrant xenobiotic metabolism, stemming from the close interactions and common regulators between lipid and drug disposition pathways [3]. Moreover, a majority of these SNPs are associated with a wide array of metabolic phenotypes highlighting the role played by genetic heterogeneity in multiple pathways in NAFLD pathogenesis, thus eliciting differences in phenotypes amongst NAFLD patients and determining the proportion of patients progressing into NASH and fibrotic stages [4]. Recent studies have also identified novel biological functions and pathways associated with NAFLD. Canonical pathway analysis has indicated an upregulation of the hepatic fibrosis pathway (*COL1A1, IL10, IGFBP3*) and a downregulation of the endoplasmic reticulum stress and protein ubiquitination pathways (*HSPA5, USP25*) along with highlighting cell development, morphology, cell movement, cell death and antigen presentation as the major biological functions associated with changes in gene expression in morbidly obese patients with NAFLD [5]. With perturbations in numerous pathways, NAFLD may be more accurately described as a 'network disease' rather than arising due to the generalised 3-hit hypothesis, thus requiring global holistic approaches to understand the disease state [6-8].

Systems biology provides global and integrative approaches to investigate complex 'network diseases' that are characterised by multiple causal mechanisms and disease outcomes. Firstly, with the advent of high-throughput data generation, it allows the correlation of vast amounts of data types. A combination of computational and systems biology provides a powerful tool to enhance the analytical abilities of experimental biology [9] and has also been utilised to identify candidate genes associated with NAFLD [10] and common disease-associated modules between NAFLD and alcoholic fatty liver disease (ALD) [11]. Secondly, it provides an opportunity to systematically model and simulate non-linear biological systems. Hence, rapid and inexpensive *in silico* simulations to test various hypotheses prior to experimentation can economise time and resource utility. Thirdly, network validation and analysis provides intuition for unidentified and uncharacterised regulations in known biological pathways that may not have been realised otherwise [12, 13].

Numerous hepatic pathway-specific models have been generated for the cholesterol synthesis pathway [14, 15], the interaction between glucose and lipid metabolism [16, 17], hepatic mitochondrial functions [18] etc. that have identified key local behavioural mechanisms in these pathways and their regulation. The HepatoNet1 is a dynamic network focusing on molecular interactions within a single hepatocyte cell [19]. It includes 777 metabolites, 2539 reactions and 1466 transport reactions. This tissue-specific genome-scale model of the hepatocyte, validated for several metabolic objectives based on the criterion for metabolic flux minimization [20], provides a comprehensive functional model of hepatocyte metabolism. Similarly, an extensive scientific venture 'The Virtual Liver Network' aims to build a multi-scale organ model of the liver with spatial and temporal attributes [21]. While these hepatocyte and liver models have immense potential to investigate liver- specific functions, they may be less informative to study the aetiology of complex systemic diseases such as NAFLD, which is characterised by deregulations in multiple tissues [22]. Hence, systemic modelling of the interaction of the liver with peripheral tissues, as described in the article, will provide greater insight into potential mechanisms of NAFLD pathogenesis.

In this article we describe a mouse-specific multi-pathway and multi-tissue model, SteatoNet (**Steato**sis **Net**work), based on ordinary differential equations (ODEs) to study the network properties of hepatic steatosis, the initial stage of NAFLD pathogenesis. The choice of mouse as the model organism, as opposed to the focus on human metabolism in the HepatoNet1 and Virtual Liver projects, provides a unique model to analyze pre-clinical experimental data. Moreover, a majority of the analyses conducted in relation to NAFLD and the regulatory mechanisms of metabolic pathways have been implemented in mouse models due to their ease of manipulation, tractability and the availability of study samples. We implement a steady-state based analysis of ODEs that does not require kinetic parameters and can uniquely identify a solution by defining the reversibility of reactions, the distribution of fluxes at pathway branch-points and the influx into the network at initial steady-state. It should however be noted here that parameters estimated from this type of analysis are semi-quantitative and provide insight into global system behaviour rather than accurate predictive values. A key feature of biological systems is the robustness displayed by pathways and the cumulative effect of regulatory mechanisms are central to maintaining this property [23]. Hence, regulatory mechanisms at both the transcriptional and post-translational level that have been associated with mouse metabolic pathways and proved experimentally to-date have been incorporated in the SteatoNet. Moreover, we mathematically show the dependence of metabolite concentration on the initial steady-state metabolic flux distribution and further conduct a sensitivity analysis on SteatoNet to identify flux distribution branch-points in various pathways that directly influence hepatic triglyceride concentration. Results from this analysis support the hypothesis of network perturbations in NAFLD pathogenesis and the importance of wide-scale modelling to understand complex biological behaviours and diseases.

## 2. METHODS

### 2.1 Dynamic modelling and steady state analysis

A systems biology library of object classes based on ordinary differential equations (ODEs) corresponding to biological pathway entities was generated for network construction, which includes the basic model generation objects such as enzyme, metabolite, non-enzymatic protein, mRNA, gene, flux sources and positive and negative regulatory objects for gene expression and protein regulation. The modelling method and model analysis in steady-state utilised has been detailed previously [15]. The reaction dynamics (Figure 1) are described by four reactions in an extension to the Michaelis-Menten model of enzymatic kinetics:

$$\frac{dS}{dt} = \phi_I (1-f) - k_C.E.S + k_{CR}.C \qquad (1)$$

$$\frac{dC}{dt} = k_C.E.S + k_{PR}.E.P - k_P.C - k_{CR}.C \qquad (2)$$

$$\frac{dP}{dt} = k_P.C - \phi_O - k_{PR}.E.P \qquad (3)$$

$$\frac{dE}{dt} = \phi_{EI} + k_P.C + k_{CR}.C - k_C.E.S - k_{PR}.E.P - \phi_{EO} \qquad (4)$$

where *S*,*E*,*C* and *P* denote the concentrations of the substrate, enzyme, substrate-enzyme complex and product respectively, $k_C$ denotes the rate constant of complex formation, $k_P$ denotes the rate constant of product formation, $k_{CR}$ denotes the rate constant of complex dissociation into the enzyme and substrate and $k_{PR}$ denotes the rate constant of the product reversibility to complex. $\phi_I$ corresponds to the substrate influx into the reaction, $\phi_O$ corresponds to the product efflux of the reaction, $\phi_{EI}$ corresponds to the influx of enzyme into the reaction, $\phi_{EO}$ corresponds to the degradation of enzyme and *f* denotes the proportion of the total substrate influx into alternative pathways.

The concentrations of *S*, *E*, *C* and *P* at steady-state are represented by $S_{ss}$, $E_{ss}$, $C_{ss}$ and $P_{ss}$. These variables can be converted into dimensionless quantities by normalising them with their steady-state counterparts. Thus,

$$S_N = \frac{S}{S_{ss}}, \qquad E_N = \frac{E}{E_{ss}}, \qquad P_N = \frac{P}{P_{ss}} \qquad (5)$$

In an extension to the method described in Belič *et al* [15], an additional steady-state ratio is described to define the relative concentration of free and bound enzyme:

$$w = \frac{C_{ss}}{E_{ss}} \qquad (6)$$

Thus, the normalised value of the complex is determined in terms of the free enzyme concentration at steady-state rather than the steady-state complex concentration. Hence,

$$C_N = \frac{C}{E_{ss}} = \frac{w.C}{C_{ss}} \qquad (7)$$

Similar to derivations in Belič *et al*, the steady state normalised values $S_N, E_N$ and $P_N$ are equal to 1 and according to the relation described above, $C_N = w$. The rate constants are now described with the incorporation of $w$:

$$k_{PN} = \frac{\phi_I(1-f)}{w(1-r)} \quad (8)$$

$$k_{CRN} = \frac{r.\phi_I(1-f)}{w(1-r)} \quad (9)$$

$$k_{CN} = \frac{\phi_I(1-f)}{(1-r)} \quad (10)$$

$$k_{PRN} = \frac{r.\phi_I(1-f)}{(1-r)} \quad (11)$$

Thus, all the model parameters can be uniquely calculated with the knowledge of the reversibility of the reaction *r*, the distribution of the influx *f* into alternative pathways, the total influx $\phi_I$ and the ratio between the bound and free enzyme, *w*. To derive the relations of the parameters at the new steady-state, $S_N{}^*, E_N{}^*$, $C_N{}^*$, $P_N{}^*$ and $f^*$, at which the system settles in the event of a disturbance, we substitute these parameters into the steady-state form of equations 1 and 2:

$$0 = \phi_I{}^*(1-f^*) - k_{CN}.E_N{}^*.S_N{}^* + k_{CRN}.C_N^* \quad (12)$$

$$0 = k_{CN}.E_N^*.S_N^* + k_{PRN}.E_N^*.P_N^* - k_{PN}.C_N^* - k_{CRN}.C_N^* \quad (13)$$

Solving for new steady-state concentration of the enzyme, substrate and product:

$$S_N{}^* = r^2.P_N{}^* + \frac{(1-f^*)(1-r^2)}{(1-f)E_N{}^*} \quad (14)$$

$$E_N{}^* = \frac{(1-f^*)(1-r^2)}{(1-f)(S_N{}^* - r^2 . P_N{}^*)} \quad (15)$$

$$P_N{}^* = \frac{1}{r^2}\left[S_N{}^* - \frac{(1-f^*)(1-r^2)}{(1-f)E_N{}^*}\right] \quad (16)$$

Equations 14-16 illustrates that apart from the classical interdependence between substrate, product and enzyme concentration, as stated by the Michaelis-Menten relations, these concentrations are also dependent on the distribution of the influx into alternative pathway branches and the reversibility of the reaction, however, they are not directly dependent on the absolute value of the total metabolic flux. The reversibility of a reaction, $r$, is an inherent property of the enzyme and the equilibrium constant between the species involved in the reaction. Although, the squared form of $r$ diminishes its influence on the concentration of the substrate, enzyme or product, it accounts for the thermodynamic constraints related to Gibb's free energy in the reactions. The flux distribution, $f^*$, in metabolic pathways in the event of a disturbance, on the other hand, is a parameter that adapts itself depending on substrate influx into the network, the demand of a product in downstream pathways, alterations in enzyme concentration and also depends on the initial flux distribution in the network, $f$.

## 2.2 Gene expression regulation

In addition to the biological entities involved directly in metabolic reactions, feedback regulation was also incorporated into SteatoNet. mRNA transcription was modelled as a sigmoid function with separate objects classes specified for positive and negative expression regulation. Negative expression control is described by the following equation:

$$\frac{dQ_{mRNA}}{dt} = \frac{Q_{max}\,\phi_0}{Q_{max} - 1 + Q_c} - k_d Q_{mRNA} \quad (17)$$

where $Q_{mRNA}$ represents relative concentration of mRNA, $Q_{max}$ is the maximum relative mRNA expression, $\phi_0$ is the transcription flux, $Q_c$ represents the concentration of the controlling molecule and $k_d$ depicts the rate of mRNA degradation. Similarly, positive expression control is described by the following equation:

$$\frac{dQ_{mRNA}}{dt} = \frac{Q_{max}\,\phi_0 Q_c (Q_{c_{max}} - 1)}{Q_{c_{max}}(Q_{max} - 1) + (Q_{c_{max}} - Q_{max})Q_c} - k_d Q_{mRNA} \quad (18)$$

where $Q_{c_{max}}$ represents the maximum concentration of the regulator that results in the maximum fold-change in mRNA expression ($Q_{max}$). The generation of protein is described as a linear relation between the relative concentration of mRNA and protein/enzyme quantity ($Q_P$):

$$\frac{dQ_P}{dt} = k_t Q_{mRNA} - k_d Q_P \quad (19)$$

where $k_t$ represents the rate constant of mRNA translation and $k_d$ is the rate constant of protein degradation. The quantity of the controlling molecule ($Q_C$) is usually controlled by activation or deactivation of the molecule by some metabolite quantity ($Q_M$), which is described in a linear manner as well. The rate equation for protein activation is:

$$\frac{dQ_{CA}}{dt} = k_a Q_{CI} * Q_M - k_d Q_{CA} \quad (20)$$

Whereas, the rate equation for protein inactivation is:

$$\frac{dQ_{CI}}{dt} = k_i Q_{CA} * Q_M - k_d Q_{CI} \quad (21)$$

where $Q_{CI}$ and $Q_{CA}$ represent the pool of inactive and active protein, respectively, $k_i$ and $k_a$ are factors describing the activation or inhibition of the protein and the $k_d$ term in equations 20 and 21 accounts for the rate of protein degradation. Although the linear representation of translation and post-translational protein regulation provides a simplified depiction of the actual process; it sufficiently represents regulatory mechanism in biology and is several fold more informative compared to models without any feedback control.

**2.3 SteatoNet construction**

An object-oriented modelling and simulation programme, Dymola (Version 7.4, Dassault Systems, Lund, Sweden), which is based on the open Modelica language, was utilized to generate SteatoNet, a dynamic non-linear model focussing on mouse-specific lipid and glucose metabolism, by compilation of reaction objects, governed by the system of equations as described in [15] and in the previous sections. Thus, a closed network of multi-reaction pathways is formed by linking reactions to each other and to regulatory objects with connectors (Figure 2). The pathways included in SteatoNet are based on mouse-specific biological pathway evidence obtained from Kyoto Encyclopaedia of Genes and Genomes (KEGG, http://www.genome.jp/kegg/) and the Reactome (www.reactome.org) databases for glucose, lipid and amino acid metabolism within the liver, adipose tissue, peripheral tissue, pancreas and macrophages. The regulation of the network at the transcriptional and post-translational level has been incorporated based on manual literature (n>500) searches.

### 2.4 Assignment Of Metabolic Flux Distributions

Considering the complex interconnected topology of biological systems, the presence of pathway branching at several points in the network is a common occurrence. The distribution of the metabolic flux, *f*, in each of the pathway branches is an independent parameter that must be specified into the model by additional equations that set the initial ratio of flux distribution from the parent pathway into each branch. Vo *et al* [24] generated a comprehensive flux network based on isotopomer tracer analysis in HepG2 cells, a hepatic carcinoma cell line. The estimated reaction fluxes from this study were utilised to approximate the value of *f* at various branch-points within SteatoNet. The choice of utilising flux analysis data from a human cell line compared to a mouse cell line was governed by the lack of comprehensive isotopomer analysis on mouse hepatocyte/liver in a normal physiological state and including a broad range of metabolic pathways. Moreover, based on flux variability analysis, Sigurdsson *et al* [25] indicated the highest degree of flux similarity between mouse and human metabolism compared to other mammals. Thus, it can be argued that the utilization of flux estimates from HepG2 cells may not significantly affect the species-specificity of the SteatoNet. Furthermore, the flux estimates from the Vo *et al* study were utilised for only approximating *f* at branch-points within SteatoNet in order to gauge a physiological value for parameter assignment, while taking into consideration that *in vitro* isotopomer studies are subject to large variances as a result of differing cell culture conditions, sampling, pre-analytical processing etc [26] and do not realistically depict organism-level flux distributions resulting from inter-tissue interactions [27]. The flux distribution proportion was calculated by summing the total flux at the branch-point and determining the proportion entering each branch as a fraction of the total flux. The choice of tracers in the Vo *et al* study and the utilization of cell lines prevented the identification of flux distributions in the lipid metabolism pathway and metabolite transport and distributions amongst different tissue types. Thus, at these branch-points with uncertainty in the flux distribution parameters, *f* was assigned an arbitrary value that resulted in stable model simulations. For several 'stiff' branch-points, the value of *f* was not completely arbitrary as these focal points can tolerate only a low range of flux distributions. Supplementary table 1 provides a complete list of the flux distribution proportions assigned to pathway branch-points in SteatoNet. It must be highlighted here that due to the semi-quantitative nature of the SteatoNet, the normalization of model parameters and the modelling goal which aims at identifying candidate mechanisms involved in NAFLD pathogenesis; the assignment of values to the parameter *f* is critical is terms of model identification rather than quantitative reproduction of hepatic function. Hence, the assignment of arbitrary flux distribution values is feasible in this case, provided that the model generates stable simulations and can simulate biological phenomena for model validation purposes.

**2.5 Sensitivity analysis**

The dependence of parameters on the model behaviour can be determined by sensitivity analysis, which is defined as the change in the model property versus the change in a parameter value [28]. Metabolic control analysis (MCA) is an extension of local sensitivity analysis to determine the extent of change in metabolic flux or other systemic properties achieved by a fractional change in enzyme activity [29]. MCA is quantified by control coefficients, which is termed as the flux control coefficient if the change in flux is considered as the model variable or as the concentration control coefficient if the change in concentration is considered as the model variable under study [28]. Relevant to this article, we can define the concentration control coefficient as the partial derivative of the change in triglyceride metabolite concentration with respect to small changes in the distribution of fluxes, $f$ , at various pathway branch-points. Thus, the sensitivity/concentration control coefficient is calculated by:

$$C_f^{TG} = \frac{d(TG)}{df} = \frac{TG^* - TG}{f^* - f} \times \frac{f}{TG} \quad (22)$$

where $C_f^{TG}$ is the concentration control coefficient of parameter $f$ with respect to hepatic triglyceride concentration $TG$ and $TG$ and $TG^*$ are the corresponding triglyceride concentrations at flux distribution values of $f$ and $f^*$. The second term in the equation, the ratio between the initial flux distribution and triglyceride concentration, is incorporated to obtain relative sensitivity coefficients that are dimensionless. Graphically, the control coefficient is the tangent to the curve describing the relation between the metabolite concentration and model parameter variation and hence, is dependent on the steady state under investigation. To determine the sensitivity of hepatic triglyceride concentration to the metabolic flux distribution, the glucose and triglyceride influx into the network was increased by 10-fold to simulate a disturbance in the system on high-glucose and high-fat diet and the distribution parameter for each branch point in the pathway model was varied by an interval of 10%. The corresponding changes in hepatic triglyceride synthesis were recorded and concentration control coefficients were calculated. In an extension of the flux distribution sensitivity analysis and equation 22, the effect of flux alterations in branch-points with high $C_f^{TG}$ on the concentration of regulatory factors in SteatoNet was determined, under similar conditions as described above, by the following equation:

$$C_f^R = \frac{d(R)}{df} = \frac{R^* - R}{f^* - f} \times \frac{f}{R} \quad (23)$$

where $C_f^R$ is the concentration control coefficient of parameter $f$ with respect to the concentration of the regulator $R$, and $R$ and $R^*$ are the corresponding regulator concentrations at flux distribution values of $f$ and $f^*$.

## 3. RESULTS AND DISCUSSION

A dynamic, semi-quantitative, multi-tissue metabolic model, the SteatoNet, focussing on the metabolic function of hepatocytes and their interaction with extra-hepatic compartments has been generated by utilising an in-house object-oriented systems biology library based on ODEs. An advantage of object-oriented modelling is the reusability of general equations and model object classes, thus avoiding tedious, error-prone manual coding methodologies for building complex systems. In addition, it also provides a user-friendly graphical interface that reduces the modelling effort. The pathways included in SteatoNet are glycolysis, gluconeogenesis, citric acid cycle, pentose phosphate pathway, *de novo* lipogenesis, β-oxidation, lipolysis, cholesterol metabolism, amino acid metabolism and ketone body synthesis. Moreover, compared to the numerous existing hepatic metabolism models, a unique highlight of the SteatoNet is the portrayal of the exchange of metabolites and regulatory molecules between the liver, adipose tissue, peripheral tissues, pancreas and macrophages, via the blood. The enzyme levels in these pathways are governed by gene expression objects, which are in turn regulated by transcription factors such as peroxisome proliferator activated receptor alpha (PPARα), PPARγ, sterol-regulatory element binding protein 1c (SREBP-1c), SREBP2, carbohydrate-response element-binding protein (ChREBP), liver X receptor (LXR), farnesoid X receptor (FXR), glucocorticoid receptor and PPARγ coactivators 1 alpha (PGC1A). Furthermore, the regulatory actions of the hormones insulin and glucagon, the adipokines leptin and adiponectin and the cytokine tumour necrosis factor alpha (TNFα) have been incorporated into the network. A total of 194 reactions involving 159 metabolites, 224 enzymes and 31 non-enzymatic regulatory proteins are represented in SteatoNet. This multi-compartmental and multi-pathway model including feedback regulation at the transcriptional and translational level provides a well-approximated portrayal of the biological system under consideration.

### 3.1 Model Validation

To determine if SteatoNet correctly depicts biological phenomena, model simulations were compared to experimental observations obtained from the literature. The model was translated by utilising the Microsoft visual studio 10.0 C compiler and the simulations were generated by utilising the default multi-step dassl solver in Dymola with a tolerance set to 1e-009. Taking into consideration the assumptions of the model, the initial normalised steady state concentration of all metabolites and enzymes has a value of 1.0 and on model perturbation the model variables are estimated based on the input values of *r, f, w* and $\Phi_I$. The simulations thus generated depict relative semi-quantitative changes in the network components in response to triggers causing a shift from the initial steady state of the model. Metabolic changes in response to fasting, the absence of stearoyl-CoA desaturase (SCD), a crucial lipogenic enzyme, and overexpression of adiponectin, an insulin-sensitising anti-inflammatory cytokine released by the adipose tissue have been well studied and were utilised to validate the model. Inconsistencies between biological observations and model simulations mainly occur due to the errors or missing objects and regulatory connections in the model. In the presence of inconsistencies, a series of simulations was implemented to identify the network objects that display erroneous behaviour and further in-depth literature searches were performed in the context of these objects to identify regulations that have been established experimentally but were absent in the network.

### 3.1.1 Fasting

The fasted or starved state is characterised by low glucose and insulin levels and increased glucagon in the blood. Glucagon activates glycogen phosphorylase for the breakdown of stored glycogen into glucose and inhibits glycogen synthase. It upregulates gluconeogenesis by increasing the expression of phosphoenolpyruvate carboxykinase (PEPCK) and downregulates fatty acid synthesis by deactivating acetyl CoA carboxylase 1 (ACC1). Glucagon also inhibits glycolysis by downregulating an activator of phosphofructokinase-1 (PFK1), a key rate-limiting enzyme in glucose metabolism [30]. Fasting has also been shown to induce peroxisome proliferator-activated receptor alpha (PPARα), a nuclear receptor that regulates mitochondrial and peroxisomal fatty acid oxidation [31]. PPARα-null mice display hypoglycaemia and an upregulation of serum free fatty acids. Additionally, fatty acid release from adipose tissue stores and oxidation is also regulated by fatty acid-induced adipose factor (FIAF), which is upregulated during fasting [32].

To simulate fasting conditions, the influx of glucose substrate into the network was reduced by 10-fold compared to the initial steady state. Figure 3 shows simulations of the levels of insulin, glucagon, blood glucose, serum fatty acids, PEPCK, ACC1, fatty acid synthase (FAS), sterol response element-binding protein- 1c (SREBP1c) and the β-oxidation enzyme, carnitine acyl transferase-1 (CPT-1) in response to the reduced glucose influx into the network. As observed from the simulations in Figure 3, the resulting downregulation of serum insulin, serum glucose, glycogen stores, lipogenic enzymes (ACC1 and FAS) and their regulator SREBP1c and increased levels of serum glucagon, serum fatty acids, the gluconeogenic enzyme, PEPCK, and the β-oxidation enzyme, CPT-1, corresponds accurately to the expected changes in the fasted state.

### 3.1.2 Stearoyl CoA knockout

Stearoyl CoA desaturase (SCD) is a rate-limiting delta-9 desaturase enzyme involved in lipogenesis [33, 34]. It catalyses the desaturation reaction to convert saturated fatty acids into monounsaturated fatty acids, in particular oleate (18:1) and palmitoleate (16:1). Monounsaturated acids serve as the key building blocks of triglycerides, cholesterol esters and membrane phospholipids [35]. The SCD-1 isoform is predominantly expressed in the liver and its expression is tightly regulated by transcription factors such as SREBP-1c [36], liver X receptor (LXR) [37] and PPARα [38], leptin [39] and estrogen [40]. *SCD -/-* mice fed with a high fat diet display a protection against diet-induced obesity and increased insulin sensitivity compared to wild-type mice due to increased fatty acid oxidation and a downregulation in lipogenesis [41]. Although a lipogenic (high-carbohydrate) diet increases the expression of genes involved in *de novo* lipogenesis, *SCD -/-* mice fed with a lipogenic diet still display low levels of triglycerides indicating the crucial role of SCD-1 in regulating triglyceride synthesis [42]. In addition, SCD-1 activity is increased in NAFLD patients [43].

To simulate the SCD knockout condition, the rate of SCD enzyme degradation was increased by 1000-fold, resulting in SCD enzyme concentration approaching 0. The lipogenic diet was simulated by increasing the glucose influx by 10-fold and the high-fat diet were simulated by increasing the triglyceride and cholesterol influx by 5-fold and 4-fold, respectively, and decreasing the glucose influx by 2.5-fold. Figure 4 illustrates the effects of SCD absence on a lipogenic diet until time $10^5$ and high-fat diet between time $10^5$ and $2x10^5$. Simulations of the response of triglycerides, fatty acid oxidation and lipogenesis enzymes to SCD knockout indicate a decrease in hepatic triglyceride accumulation and increase in the level of CPT-1 on both diets. While the lipogenic enzyme glycerol-3-phosphate acyltransferase (GPAT) and its transcriptional regulator SREBP-1c were upregulated under lipogenic diet conditions, their concentrations decreased on a high-fat diet, in concordance with experimental observations in rodent models.

### 3.1.3 Adiponectin overexpression

Adiponectin is an adipose-secreted cytokine that has a negative correlation with insulin resistance, plasma triglycerides and low-density lipoprotein (LDL) – cholesterol, hepatic fat content and progression to NASH in NAFLD patients [44-46]. It is known to affect glucose metabolism by repressing hepatic glucose production via the activation of adenosine monophosphate-activated protein kinase (AMPK) α-2 in hepatocytes [47]. The beneficial impact of adiponectin on metabolism is enforced by the ceramidase activity of its receptors, AdipoR1 and AdipoR2, to lower hepatic levels of ceramides, which are bioactive lipids implicated in insulin resistance [48]. Shetty *et al* [49] observed a 30% decrease in plasma fatty acids, an upregulation in ketone bodies and fatty acid oxidation in the adipose tissue and decreased expression of lipogenic genes on fasting in adiponectin-overexpressing transgenic mice. Whilst adiponectin has insulin-sensitizing, anti-inflammatory and antilipogenic effects, the adipokine, tumour necrosis factor alpha (TNF-α) is thought to be a suppressor of adiponectin, with adiponectin-null mice displaying high levels of TNF-α mRNA and protein levels and diet-induced insulin resistance, thus explaining the phenomenon of increased TNF-α levels in obese populations [50]. The proinflammatory and insulin-desensitising effects of TNF-α, potentially via the activation of other cytokines such as inhibitor of nuclear factor kappa-B kinase subunit beta (IKKβ) and nuclear factor kappa-B (NF-κB), is well documented [51]. Thus, a balance between adiponectin and TNF-α level may determine the susceptibility to NAFLD progression.

To simulate the overexpression of adiponectin under fasting conditions, the rate of adiponectin degradation was decreased by 10-fold, resulting in increased levels of adiponectin and the glucose substrate influx into the network was reduced by 10-fold. The levels of hepatic triglycerides, serum fatty acids, hepatic lipogenic genes, ketone bodies, adipose β-oxidation enzymes, TNF-α and ceramides were simulated in response to the overexpression of adiponectin (Figure 5). In concordance with experimental observations, increased adiponectin concentration results in downregulation of serum fatty acids, hepatic triglycerides, TNF-α, ceramides, lipogenesis mediators SREBP1c, GPAT and SCD1 and increased expression of adipose CPT-1 indicating an upregulation of β-oxidation in adipocytes. The model simulation however, did not indicate any changes in serum ketone body concentration on adiponectin overexpression.

### 3.2 Flux distribution sensitivity analysis

Metabolic flux through a pathway is a crucial parameter and its value may depend upon flux entering the network, enzyme activity or concentration and genetic variations. The interconnected nature of metabolic networks results in the frequent occurrence of branch-points in pathways resulting in the distribution of fluxes into daughter branches. Since NAFLD has been previously reported to result from increased fatty acid flux from adipose tissue [22], it may be hypothesised that additional flux distributions may be disrupted in the disease. Thus, we aimed to conduct a sensitivity analysis to identify flux branch points in the SteatoNet that display a significant influence on hepatic triglyceride accumulation.

The concentration control coefficients with respect to hepatic triglyceride, $C_f^{TG}$, of each branch-point was determined over a broad range of flux distribution percentages on system perturbation. According to their $C_f^{TG}$, the branch-points were classified as focal points with 'high sensitivity' ( $C_f^{TG}$ >1), 'moderate sensitivity' ($0.1 \ll C_f^{TG} \ll 0.99$), 'low sensitivity' ($C_f^{TG}$<0.1) or 'null sensitivity' ($C_f^{TG}$ = 0). Amongst the high sensitivity branch-points, those that displayed dynamic changes in their $C_f^{TG}$ over the varying flux range were further sub-classified as 'high dynamic sensitivity' branch-points (Figure 6). In addition, several branch-points displayed low tolerance to flux changes and were sub-classified as 'high/moderate/low sensitivity + low tolerance' branch-points (Tables 1-3). The sensitivity of transcription and regulatory factors to alterations in branch-point flux distributions was also determined. Pathway branch-points with high dynamic or high dynamic + low tolerance $C_f^{TG}$ and regulatory factors that are sensitive to alterations in flux distributions at these branch-points are illustrated in Figure 7.

### 3.2.1 Branch-points with high dynamic $C_f^{TG}$

Sensitivity analysis in SteatoNet identified the prevalence of high dynamic sensitivity branch-points in the glucose metabolism and transport, lipid transport and ketone body synthesis and transport pathways. While the tolerance to varying flux distributions at these branch-points is high, their $C_f^{TG}$ displayed significant variations over the broad flux distribution range. As observed from Figure 7, a majority of the high dynamic sensitivity branch-points (in bold) occur in the inter-tissue transport pathways, highlighting the crucial role played by metabolite transport and redistribution amongst the different tissue types.

***Lipid Metabolism***

Figure 7 illustrates an enrichment of the lipid metabolic pathways amongst the branch-points with high dynamic $C_f^{TG}$. These include activation and desaturation of adipose fatty acids and the reactions that contribute to the adipose fatty acid pool such as fatty acid lipolysis in the adipocyte and transport between the blood, adipocyte and peripheral tissues (Figure 6a and b, Figure 7); the contribution of the diet via chylomicrons and VLDL breakdown to serum fatty acids (Figure 6c, Figure 7); and the transport of serum cholesterol to the liver (reverse cholesterol transport via HDL), adipose tissue and peripheral tissues (Figure 6d and e, Figure 7). Previous studies have indicated an increased flux of adipose-derived non-esterified fatty acids that contributes to hepatic triglyceride concentration in NAFLD patients [22]. In correspondence with this observation, the high negative $C_f^{TG}$ of the desaturation and activation of adipose fatty acids and the transport reactions contributing to the adipose fatty acid pool highlighted by the SteatoNet (Figure 6a and b) indicates the negative correlation between adipose-specific lipogenesis and hepatic triglyceride concentration.

The inter-tissue distribution of cholesterol and source of accumulated hepatic cholesterol in NAFLD patients or animal models has not been clearly established.

Cholesterol accumulation, along with increased expression of low density lipoprotein receptor (LDLR) or genes in the cholesterol synthesis pathway has been observed in an obese insulin-resistant rodent model of NAFLD and NAFLD patients respectively [52, 53], which contributes to hepatic lipotoxicity and progression to NASH [52, 54]. A study in obese children indicated an increased risk of NAFLD development in cases with increased total serum cholesterol and LDL [55]. In accordance with these clinical observations, the sensitivity analysis indicated the role of the reverse cholesterol transport pathway and deregulated cholesterol distribution to the adipose and other peripheral tissues in hepatic steatosis (Figure 6d and e, Figure 7). The highly dynamic positive $C_f^{TG}$ of LDL distribution to the adipose and peripheral tissue implies that low to moderate LDL fluxes towards non-hepatic compartments does not affect hepatic triglyceride concentration significantly. However, at higher LDL flux distributions to the adipose and peripheral tissues, the positive correlation to hepatic triglyceride increases exponentially. This observation can be explained by studies that have indicated the role of LDL-cholesterol transport to the liver in ensuring triglyceride-rich VLDL secretion [56]. Thus, there appears to be a fine balance in the concentration and source of cholesterol in determining its role in sustaining VLDL secretion to prevent steatosis or causing lipotoxicity. It may also be speculated that these paradoxical roles of cholesterol establish the stage of NAFLD (steatosis or NASH). Additionally, tracer studies with labelled substrates may prove useful to determine the contribution of *de novo* cholesterol synthesis and lipoprotein-derived cholesterol in establishing the fate of cholesterol in the NAFLD disease state. These observations thus lay emphasis on the importance of cholesterol in NAFLD pathogenesis even at early stages of the disease, in addition to the currently accepted notion of cholesterol-induced lipotoxicity in triggering NASH.

### *Glucose metabolism*

The exchange of glucose between the serum and liver, the initial steps of the glycolysis pathway, the transport of glucose to the adipose tissue and the contribution of dietary glucose to serum glucose level display a high dynamic influence on hepatic triglyceride concentration (Figure 6f-h, Figure 7). High carbohydrate consumption, especially fructose, has been linked to triggering NAFLD [57, 58] as sugars provide precursors for fatty acid, triglyceride and cholesterol synthesis. Additionally, glucose metabolism has been implicated in NAFLD as type II diabetes and insulin resistance are predominant characteristics in NAFLD patients and are associated with the severity of the disease [59, 60]. Insulin plays a crucial role in glucose uptake and gluconeogenesis inhibition and additionally regulates *de novo* lipogenesis by activating SREBP1-c and thus upregulating ACC1 and FAS [61, 62]. The lipogenic role of insulin may thus be causal of the moderately positive $C_f^{TG}$ of the hepatic release of glucose into the blood (Figure 6h, Figure 7), which triggers insulin secretion. The high dynamic positive $C_f^{TG}$ of glucose transport to adipose tissue (Figure 6d, Figure 7) indicates a positive correlation between hepatic triglyceride concentration and increased glucose flux to adipose tissue, which potentially stems from the resulting upregulation of precursors available for *de novo* lipogenesis in adipocytes and hence, larger lipid depots that increase the fatty acid flux towards the liver on lipolysis. Hence, the transport of glucose and increased dietary intake may directly provide increased substrate flux for *de novo* triglyceride synthesis or/and may indirectly be involved in regulating hepatic triglyceride concentration via the action of insulin.

The high dynamic negative $C_f^{TG}$ of the generation of fructose-6-phosphate from glucose-6-phosphate (Figure 7 and Figure 6f), the first step of glycolysis, indicates a general negative correlation between hepatic triglyceride concentration and glycolysis. The negative correlation can be explained by the decreased activation of lipogenic factors ChREBP and SREBP1c by glucose [63] and insulin [64], respectively, as a result of glucose catabolism. The dynamic profile of the $C_f^{TG}$ indicates that at low flux proportions towards glycolysis, the decrease in hepatic triglyceride concentration is much more significant than at higher flux values due to the low availability of precursors for *de novo* lipogenesis as well as decreased glycerol-3-phosphate production, which is required for triglyceride synthesis. Thus, the SteatoNet indicates the beneficial effects of glucose oxidation on hepatic steatosis however; the extent of the benefit is dependent on the glycolytic flux directed towards lipogenesis. At low glycolytic flux, the role of glucose is predominantly in ATP-generating oxidation via the TCA cycle rather than *de novo* lipogenesis.

***Ketone body metabolism***

Sensitivity analysis further highlighted the high negative $C_f^{TG}$ of the synthesis and transport of ketone bodies, acetoacetate and β-hydroxybutyrate (BHB) (Figure 6i-k, Figure 7) indicating a negative correlation between hepatic triglyceride concentration and ketogenesis. In concordance with this observation, a genetic screen in a zebrafish model of hepatic steatosis highlighted the role of solute carrier family 16a, member 6 a (Slc16a6a), a transporter of BHB in hepatic accumulation of triglycerides [65]. A mutation in this transporter diverted BHB as a substrate for *de novo* lipogenesis. Serum levels of ketone bodies also indicate the extent of mitochondrial β-oxidation in the liver [66]. Thus, the synthesis and extra-hepatic transport of ketone bodies is an indicator of hepatic triglyceride concentrations, as confirmed by the sensitivity analysis in our model. However, the less prominent downregulation of hepatic triglyceride concentration at increased fluxes in ketone body synthesis and transport pathway may occur due to a decrease in mitochondrial acetyl CoA flux towards the citric acid cycle for energy release, in addition to the increased availability of precursors for triglyceride synthesis in adipose tissue that may eventually be transported to the liver.

### 3.2.2 Stiff branch-points with high $C_f^{TG}$

Several pathway branch-points displayed high, medium or low $C_f^{TG}$ but only over a limited flux distribution range. Beyond this range, the perturbed model incurred instability in the absence of attainment of a new steady state. Branch-points with high values of $C_f^{TG}$ and low flux range tolerance (Table 1, Figure 7, not in bold) are mainly present in the triglyceride metabolism pathway, which include the formation of hepatic lysophosphatidic acid (LPA), the first step in triglyceride synthesis; the breakdown of adipose triglycerides into diacylglycerol (DAG); the storage of triglycerides in adipose lipid droplets and the synthesis of adipose dihydroxyacetone phosphate (DHAP), which is a precursor for glycerol-3-phosphate involved in *de novo* lipogenesis (Figure 7). These focal points in the network may play a key initiating role in NAFLD by causing instability under conditions of even slight alterations in flux distribution. This observation is supported by several studies implicating the role of adipose triglyceride storage/hydrolysis by mediators such as perilipin [67], a lipid droplet-associated protein, adipose triglyceride lipase (ATGL), a triglyceride lipase [68, 69], comparative gene identification-58 (CGI-58), a co-activator of ATGL [67] and hormone sensitive lipase (HSL) [70] in hepatic lipid accumulation. Furthermore, ATGL, which has highest affinity for triglyceride lipolysis in the adipocyte, plays a role in hepatic endoplasmic reticulum (ER) stress, a characteristic of NAFLD [71]. The stiffness in the distribution of hepatic fatty acids to form LPA is crucial due to its role in *de novo* lipogenesis. LPA is the substrate for the lysophosphatidic acid acyltransferase (LPAT) reaction catalysed by adiponutrin/PNPLA3 [72]. A polymorphism *rs738409 C/G*, encoding the PNPLA3 I148M variant has been consistently identified in NAFLD GWAS studies in various populations [4, 73-75]. The molecular function of PNPLA3 and the effect of the *rs738409* polymorphism on it are only recently being understood [72, 76]. Li *et al* reported that overexpression of PNPLA3 I148M in the liver, but not adipocyte, results in hepatic steatosis in mice due to increased triglyceride synthesis and reduced triglyceride hydrolysis. The increase in triglyceride synthesis is potentially via the monoglyceride pathway as no accumulations in phosphatidic acid or DAG were observed. However, overexpression of wild-type PNPLA3 has no effects on hepatic fat indicating additional regulatory effects of the polymorphism apart from overexpression in triggering steatosis. The low tolerance of the branch-point determining the distribution of fatty acids to LPA formation may act as a mechanism to ensure limited availability of substrate for triglyceride synthesis and hence, a breach of this limited flux range may trigger hepatic triglyceride accumulation, especially in the presence of the *PNPLA3 I148M* polymorphism.

Several branch points with medium or low values of $C_f^{TG}$ also display low flux range tolerance in several pathways (Table 2 and Table 3). Branch-points with moderate sensitivity include the lipid metabolic pathways involving LPA formation in the adipocyte, triglyceride hydrolysis in the liver, fatty acid transport from the serum to the liver and the formation of cholesterol esters.

### 3.2.3 Regulators of branch-points with high $C_f^{TG}$

Pathway branch-points with high values of $C_f^{TG}$ were further investigated to identify molecular regulators that are sensitive to alterations in flux distributions at these branches, by calculating their concentrations control coefficients with respect to various regulatory proteins in the SteatoNet, $C_f^{R}$ (equation 23). Figure 7 illustrates the regulatory factors with high sensitivity to alterations in flux distributions at the respective branch-points. FXR, a regulator of bile acid synthesis and excretion, LXR, a sterol-sensor, and SREBP2, the key regulator of cholesterol synthesis, display global sensitivity to alterations in metabolic flux distribution at a majority of the high sensitivity pathway branches, indicating the broad role of these transcription factors that predominantly have a common function in cholesterol metabolism. FXR is repressed by the transcription factor Yin Yang 1 resulting in hepatic steatosis in high-fat diet-induced obese mice [77] and has been identified as a target for NAFLD treatment [78]. Moreover, NAFLD patients display low levels of FXR expression along with increased expression of LXR and SREBP1c, which contributes to increased hepatic triglyceride synthesis [79]. Excess cholesterol accumulation in the liver triggers the activation of LXR for cholesterol export and inhibits SREBP2-mediated *de novo* cholesterol synthesis by an auto-inhibitory mechanism [80, 81]. Thus, the global sensitivity of FXR, LXR and SREBP2, as observed in the SteatoNet, further emphasises on the role of cholesterol in maintaining lipid homeostasis in the liver.

Interestingly, a majority of the transcription factors and metabolic regulators displayed sensitivity to alterations in the flux distribution in the glucose transport pathway (Figure 7) indicating the global effect of glucose on metabolic regulation. Furthermore, it can be observed that transcriptional factors that are not directly involved in regulating a particular pathway may still be sensitive to the distribution of fluxes into the pathway (Figure 7) e.g. the sensitivity of FXR and SREBP2 to branch-points in the early hepatic glucose metabolism pathway. This can be explained by the largely interconnected nature of metabolic networks, whereby products generated by one pathway are utilised as substrates by another pathway. Thus, the inter-dependency of pathways may simultaneously be controlled by a balance in the levels of various transcriptional factors, providing further evidence of the network-wide disturbances in NAFLD-related steatosis.

## 4. CONCLUSIONS

As rightly suggested by Lanpher *et al* [82], most complex diseases can be described as less severe cases of Mendelian inborn errors of metabolism (IEM) in which several pathways are subtly affected. The cumulative effect of genomic variations and environmental/dietary factors results in altered metabolic flux distributions and hence, a broad spectrum of disease phenotypes. Thus, utilising a holistic approach, in addition to traditional reductionist methods, to study disease-related networks, termed as systems pathobiology, will provide a clearer portrait of the systemic deregulations in complex diseases [83]. It will also enable drug development strategies to be redirected towards systemic medicine and multi-targeting approaches to identify appropriate treatment regimen. In this article we provide evidence that NAFLD is not trivially the hepatic manifestation of the metabolic syndrome, as commonly described according to Oslerian theory [84] that focuses specifically on the dysfunctional organ in a disease, but instead arises as a result of network-wide perturbations at the organism-level. While experimental evidence has long before indicated the systemic nature of NAFLD pathogenesis involving the role of the adipose tissue [85, 86], skeletal muscles [87, 88], intestine [89] and the heart [90, 91], there has been a lack of directed efforts to study NAFLD as a systemic condition.

The validated multi-tissue metabolic network, SteatoNet, presented here provides a cohesive illustration of the cascade of disruptions caused by altered flux distributions that may have been triggered in a separate tissue such as the adipose, ultimately resulting in hepatic steatosis. SteatoNet highlights the ability of models to sufficiently represent biological behaviour in steady-state analysis in the absence of kinetic parameters. Parameter estimation is a major hurdle in modelling biological networks as the extrapolation of kinetic parameters from isolated *in vitro* studies to multi-pathway networks has several drawbacks and need careful interpretation due to the presence of inconsistencies in kinetic measurements and multiple isoforms of enzymes, which are tissue- or substrate- specific. Fitting models based on optimizing model behaviour, objective functions and experimental data leads to large inconsistencies in parameter values and hence, unless very accurately measured, kinetic parameters cannot be constrained to generate useful predictions [92]. The normalization of variable parameters in SteatoNet provides a semi-quantitative description of the model variables, thus curbing the need for specifically constraining parameters using kinetic data that are at present poorly documented, display variability and are measured *in vitro* under sub-optimal conditions that are incompatible to molecular behaviour *in vivo*. Model validation and the identification of flux disturbances by sensitivity analysis that have been previously proven experimentally in NAFLD patients/models indicate that accurate model parameter estimation is not critical for models investigating systemic steady-state behaviour to identify candidates for complex diseases. In cases where the scope of the model requires the accurate dynamic investigation of specific variables, especially to identify fine time-dependent alterations in molecular dynamics, parameter estimation is indisputably a crucial step. As the scope of SteatoNet was aimed at providing an overview of systemic steady-state behaviour to identify pivotal focal-points of system instability that may be causal of NAFLD pathogenesis, the simplification of kinetic behaviour does not affect the performance of the model. While simplifying reaction-specific information, we stress on the importance of expansiveness and structural accuracy in models of biological systems. In spite of the success of dynamic modelling efforts, a majority of the models generated for eukaryotic organisms portray a simplified picture of molecular interactions. While a simplification of intra-pathway complexity is advantageous in terms of model analysis, it leads to a loss of information if inter-pathway and tissue-tissue interactions are over-simplified. Palsson *et al* proved the importance of the completeness in model structural complexity by comparing models with different complexity levels for interpreting results from model sensitivity analysis [93]. Since, the physiological state of organisms is a result of complex system-wide interactions, analysing individual pathways in isolation leads to loss of information and inaccurate interpretations. Such an approach is a fundamental primary step in systems pathobiology by analysing the broad scheme of the ‘candidate’ disease network and then channelling efforts towards specific molecular interactions.

An additional dominant feature of the SteatoNet is the portrayal of feedback regulatory mechanisms at the transcriptional and post-translational level. Robustness is an inherent property of biological systems resulting from complex feedback control mechanisms elicited by a multitude of regulatory factors [23]. In the presence of system perturbations, robustness enables systems operating at an equilibrium state to either obtain a different nominal steady state or restore the initial equilibrium. Despite the simplified linear depiction of feedback regulation in SteatoNet, the inclusion of regulatory mechanisms is of fundamental nature and beneficial in maintaining network robustness and concurring effectively with biological behaviour.

Here, we have provided a comprehensive sensitivity analysis highlighting the effects of altered flux distributions in metabolic pathways on hepatic steatosis. The critical dependence of hepatic triglyceride concentration on inter-tissue transport reactions highlights the multi-tissue nature of NAFLD and thus, the disruption in the homeostatic balance in the distribution of numerous metabolites is at least in part causal of the disease state. Moreover, the analysis provides a hypothesis for the chain of events that trigger the disease whereby the accumulation of hepatic triglycerides is initially triggered due to a deregulation in stiff pathway branch-points with low tolerance to flux alterations. The function of these stiff focal points in the early steps of hepatic lipogenesis and in the adipose compartment in maintaining a balance in triglyceride synthesis, storage and hydrolysis may indicate their initial role in triggering hepatic steatosis. The “threshold effect” displayed by the stiff branch-points beyond which the altered metabolic fluxes result in disease initiation may be specific to individuals and potentially explain the variability in NAFLD pathogenesis, especially in the case of the stiff LPAT reaction catalysed by adiponutrin, whose PNPLA3 I148M variant is commonly observed in diverse NAFLD populations. Instability in these low tolerance reactions may elicit functional changes in regulatory factors such as FXR and LXR, which in turn activate a cascade of alterations in numerous pathways that are sensitive to these factors, thus resulting in a pathological state. In addition to confirming hepatic triglyceride sensitivity to previously identified deregulations in NAFLD, the model has allowed the identification of candidate mechanisms such as cholesterol, glucose and ketone body transport and metabolism and regulatory functions of FXR, LXR and SREBP-2 that require experimental focus in the future. However, the question still remains on potential mechanisms that result in altered flux distributions. As stated by Magnuson *et al*, the regulation of metabolic flux is a complex phenomenon that cannot be pinpointed to a single mechanism, as metabolic pathways are intricately connected, involve multiple enzyme isoforms with different substrates that are regulated by numerous factors at the transcriptional and translational level [94]. Moreover, the inconsistency between the high prevalence of NAFLD and the small number of genetic variants identified in association with the disease indicates the critical role of environmental and dietary factors in initiating hepatic steatosis. Excessive dietary lipid or carbohydrate intake, and hence, overproduction of metabolic substrates may result in enzyme saturation, forcing a shift of metabolic flux into alternative pathways. In addition, dietary factors cause diverse changes in the metabolic status of an organism by triggering the activation or inhibition of regulatory factors such as hormones, cytokines and nuclear receptors [95, 96]. Individuals with genetic variations or inborn errors in metabolism affecting the activity or concentration of enzymes or regulatory factors may be predisposed to hepatic steatosis as a result of inherent deregulations in flux distributions, with further adverse effects in the absence of suitable dietary measures.

In spite of the significance and informative nature of the SteatoNet in identifying candidate mechanisms/mediators in NAFLD pathogenesis and generating hypotheses for experimental validation, we are aware of drawbacks such as the model’s lack of quantitative prediction capacity. The problem of parameter estimation in complex models is a major limiting factor for accurate reconstructions of biological pathways and thus, modelling single cell or multi-compartmental model systems often involves assumptions to simplify complex biological phenomena. Hence, models are by no means a complete picture of the system under study and results from model analysis need to be interpreted with caution. However, the semi-quantitative accuracy obtained from the dynamic non-linear system presented here indicates the strength in the structural properties of models rather than accurate parameter estimation, which requires extensive time-course experiments and measurements of a large number of variables. Additionally, the one-at-a-time method of local sensitivity analysis implemented in this article estimates sensitivity to only single parameters rather than the interaction between several parameters, which requires increased computational time due to the large number of simulations that need to be generated to test variations in numerous parameters simultaneously. Using variance-based methods that account for multi-parameter interactions is a topic of future interest, to determine if multiple flux alterations influence the degree of hepatic steatosis. Future work will involve the utilisation of the model to aid in experimental design and establish it as an integrative platform to analyze different data types. The non-specific nature of the model also enables its utility to investigate a wide-range of biological phenomena and other metabolic disorders. While the scope of this study was limited to identifying the flux branch-points that potentially trigger hepatic steatosis, the first stage of NAFLD pathogenesis, it may be interesting to determine factors that affect the progression of steatosis into NASH and hepatocellular carcinoma by analyzing the sensitivity of inflammatory and oncogenic mediators to deregulated lipid homeostasis. The inclusion of interactions with other relevant compartments such as the gut and heart may also highlight additional mechanisms involved in NAFLD.

In summary, SteatoNet provides a unique multi-tissue, dynamic, semi-quantitative platform for *in silico* ‘experimentation’, in order to investigate complex metabolic disorders such as NAFLD. The strength of SteatoNet lies in its structural accuracy, robustness, expansiveness and the inclusion of regulatory feedback mechanisms at various hierarchical levels controlling metabolic networks that allows systemic behavioural analysis in the absence of estimated kinetic parameters. The network characteristic of NAFLD pathogenesis triggered due to alterations in flux distributions at low-threshold focal points in the network, which cascades the perturbations to interacting pathways (cholesterol, glucose and ketone body transport) and regulatory factors directly influencing hepatic triglyceride concentration, a hypothesis suggested by SteatoNet analysis, requires experimental validation.

**ACKNOWLEDGEMENTS**

This work was funded by the Marie Curie FP7 initial training network “FightingDrugFailure” (Grant Agreement Number 238132).

| PATHWAY BRANCH | FLUX RANGE | SENSITIVITY |
|---|---|---|
| **FA (+ Gly-3-P) to LPA** | Upto 10% of total flux into fatty acids | 2.098 |
| **TG storage in $LD_A$** | 20-30% of total flux into adipocyte triglycerides | 1.224 |
| **$TG_A$ to $DAG_A$** | 50-60% of total flux into adipose triglycerides | 2.781 |

**Table I. Pathway branch-points with high sensitivity and low flux range tolerance.** FA- Fatty acids, Gly-3-P- Glycero-3-phosphate, LPA- Lysophosphatidic acid, TG- Triglycerides, LD- Lipid droplets, $_A$- adipose compartment.

| PATHWAY BRANCH | FLUX RANGE | SENSITIVITY |
|---|---|---|
| **G-6-P to G-1-P** | Upto 40% of total flux into glucose | -0.6 to -0.2 |
| **Cholesterol + SFA CoA to Cholesterol esters** | Upto 10% of total flux into saturated fatty acyl CoA | -0.2 |
| **$FA_A$ (+ Gly-3-P) to $LPA_A$** | Upto 30% of total flux into fatty acyl $CoA_A$ | -0.5 to -0.14 |
| **$FA_B$ to $MUFA_L$** | Upto 30% of total flux into serum fatty acids | 0.14 to 0.36 |
| **$TG_L$ to $DAG_L$** | Upto 40% of total flux into liver triglycerides | 0.3 |

**Table II. Pathway branch-points with medium sensitivity and low flux range tolerance.** G-6-P- Glucose-6-phosphate, G-1-P- Glucose-1-phosphate, SFA- Saturated fatty acids, FA- Fatty acids, Gly-3-P- Glycero-3-phosphate, LPA- Lysophosphatidic acid, MUFA- Monounsaturated fatty acids, TG- Triglycerides, DAG- Diacylglycerol, $_A$- adipose, $_B$- blood/serum, $_L$-liver.

| PATHWAY BRANCH | FLUX RANGE | SENSITIVITY |
|---|---|---|
| **G-6-P to Ribulose-5-P** | Upto 30% of total flux into glucose | -0.08 to -0.05 |
| **Serine to Glycine** | Upto 30% of total flux into serine | -0.00015 to 0.00056 |
| **Serine+Homocysteine to Cystathionine** | Upto 30% of total flux into serine | 0.00235 to 0.00475 |
| **Glutamate to Glutamic semialdehyde** | Upto 20% of total flux into glutamate | 0.0032 |
| **DAG+Choline to Phosphatidylcholine** | Upto 40% of total flux in DAG | 0.015 to 0.031 |
| **Cholesterol + USFA CoA to Cholesterol esters** | Upto10% of total flux into USFA CoA | -0.02 |
| **$\text{Cholesterol}_A$ synthesis** | Upto 20% of total flux into acetyl CoA adipocyte | 0.00014 |
| **Acetyl $\text{CoA}_T$ to Malonyl $\text{CoA}_T$** | Upto 30% of total flux into acetyl CoA tissue | -5.12E-06 to -7.89E-06 |
| **$\text{Cholesterol}_T$ synthesis** | Upto 20% of total flux into acetyl CoA tissue | 0.000102 |
| **DHAP to Gly-3-P** | Upto 20% of total flux into DHAP | -0.07 |
| **DAG to MAG** | Upto 30% of total flux into DAG | -0.1 to -0.04 |

**Table III. Pathway branch-points with low sensitivity and low flux range tolerance.** G-6-P- Glucose-6-phosphate, Ribulose-5-P- Ribulose-5-phosphate, DAG- Diacylglycerol, USFA- Unsaturated fatty acids, DHAP- Dihydroxyacetone phosphate, Gly-3-P- Glycero-3-phosphate, MAG- Monoacylglycerol, $_T$- peripheral tissues.

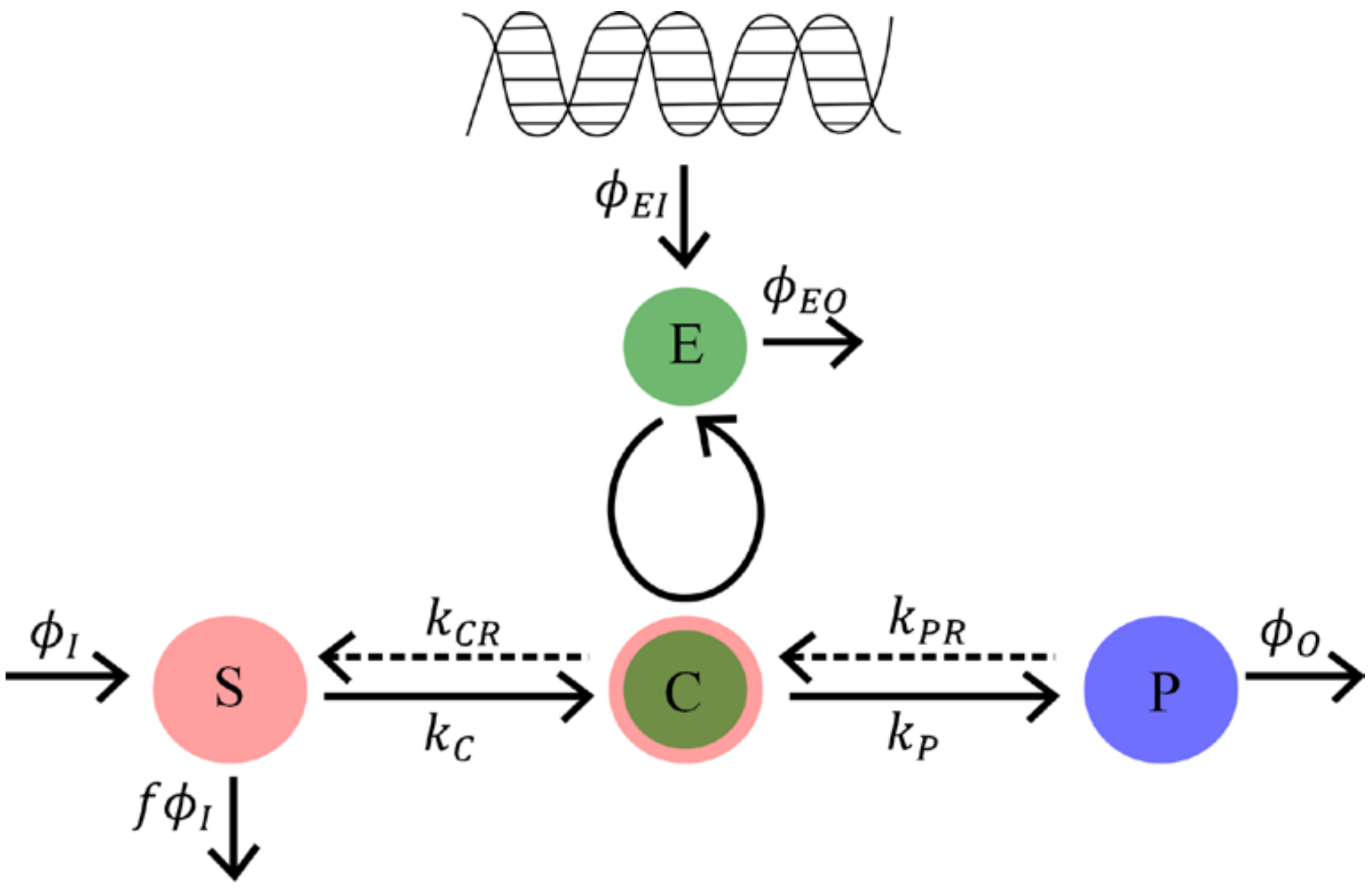

**Figure 1. Dynamics of an enzymatic reaction according to the Michaelis-Menten kinetic formalism.** *S,E,C* and *P* denote the concentrations of the substrate, enzyme, substrate-enzyme complex and product respectively, $k_C$ and $k_P$denote the rate constant of complex formation and product formation respectively, $k_{CR}$ and $k_{PR}$ denote the reverse reaction rate constant of complex dissociation into the enzyme and substrate and product reversibility to complex respectively. $\Phi_I$ corresponds to the substrate influx into the reaction, $\Phi_O$ corresponds to the product efflux of the reaction, $\Phi_{EI}$ corresponds to the influx of enzyme into the reaction, $\Phi_{EO}$ corresponds to the degradation of enzyme and *f* denotes the proportion of the total substrate influx into alternative pathways.

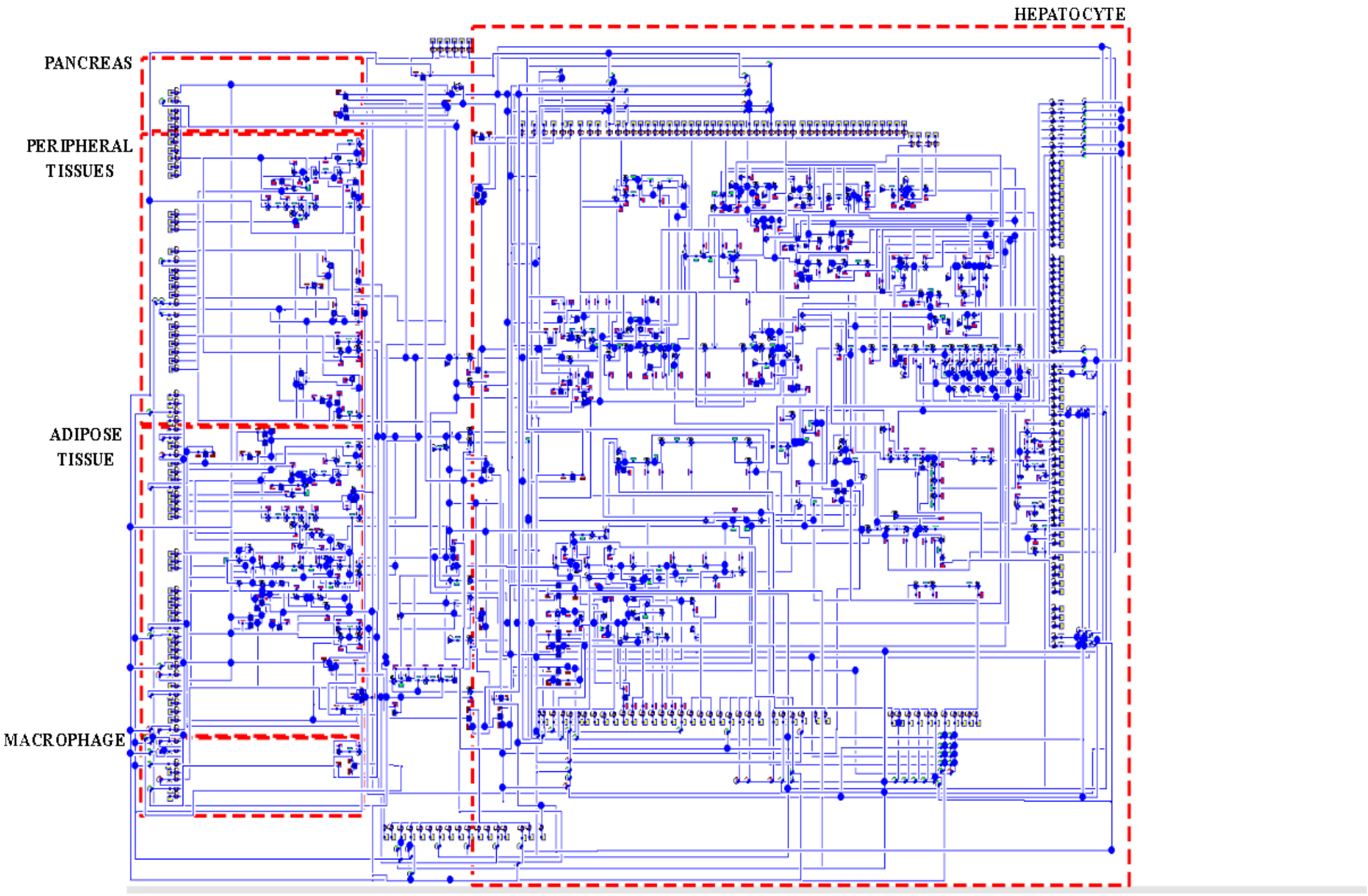

**Figure 2. SteatoNet metabolic network.** The multi-compartmental network was generated by systematically linking objects representing biological entities (e.g. metabolites, proteins, gene expression etc.) from the systems biology library, based on pathway evidence from KEGG, the Reactome and literature searches. The key metabolic pathways including glucose, fatty acid, cholesterol and amino acid metabolism and their regulation by hormonal, adipokine and transcriptional and post-translational regulatory factors are represented in the hepatic, adipose, macrophage, peripheral tissue and pancreatic compartments with inter-tissue connectivity via the blood. The SteatoNet consists of 194 reactions involving 159 metabolites, 224 enzymes and 31 non-enzymatic regulatory proteins.

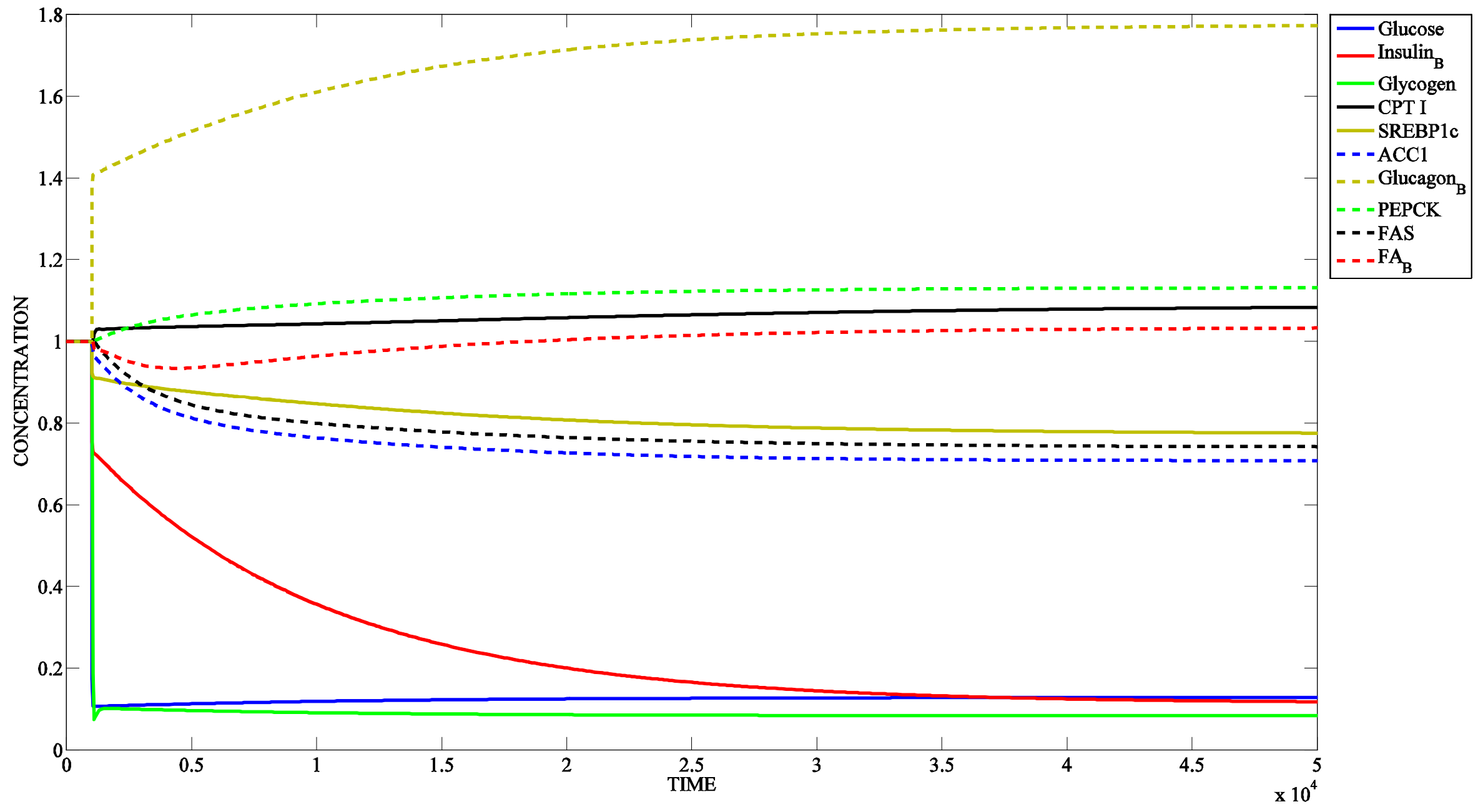

**Figure 3. Model validation of fasting conditions.** Fasting conditions were simulated by reducing the influx of the glucose into the network by 10-fold. In accordance with biological observations, model simulations indicate a downregulation of serum insulin, serum glucose, glycogen stores, lipogenic enzymes (ACC1, FAS and SREBP1c) and increased levels of serum glucagon, PEPCK, serum fatty acids ($FA_B$) and the β-oxidation enzyme, CPT-1 in response to the change in glucose influx into the network.

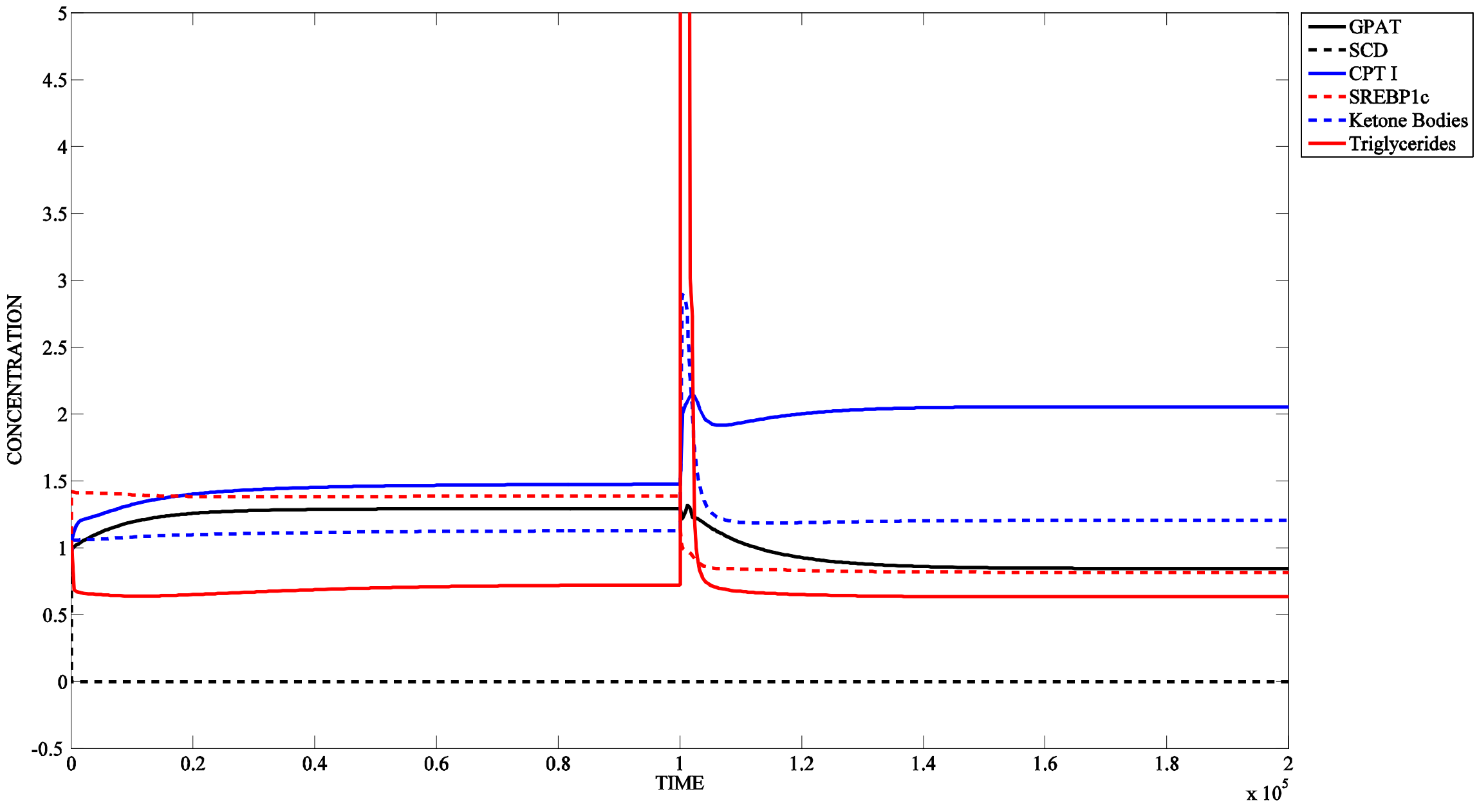

**Figure 4. Model validation of SCD knockout conditions on a lipogenic and high-fat diet.** The SCD knockout condition was simulated by increasing the rate of SCD enzyme degradation by 1000-fold, resulting in SCD enzyme levels approaching 0. The lipogenic diet was simulated by increasing the glucose network influx by 10-fold (until time $1x10^5$) and the high fat diet was simulated by increasing the chylomicron and cholesterol network influx by 5-fold and 4-fold (between time $1x10^5$ and $2x10^5$), respectively and decreasing the glucose influx by 2.5-fold. The absence of SCD resulted in a decrease in hepatic triglyceride accumulation and increase in the levels of CPT 1 and ketone bodies on both diets. The lipogenic diet triggered an upregulation of FAS and GPAT and their transcriptional regulator SREBP-1c, whereas these proteins were downregulated on the high fat diet.

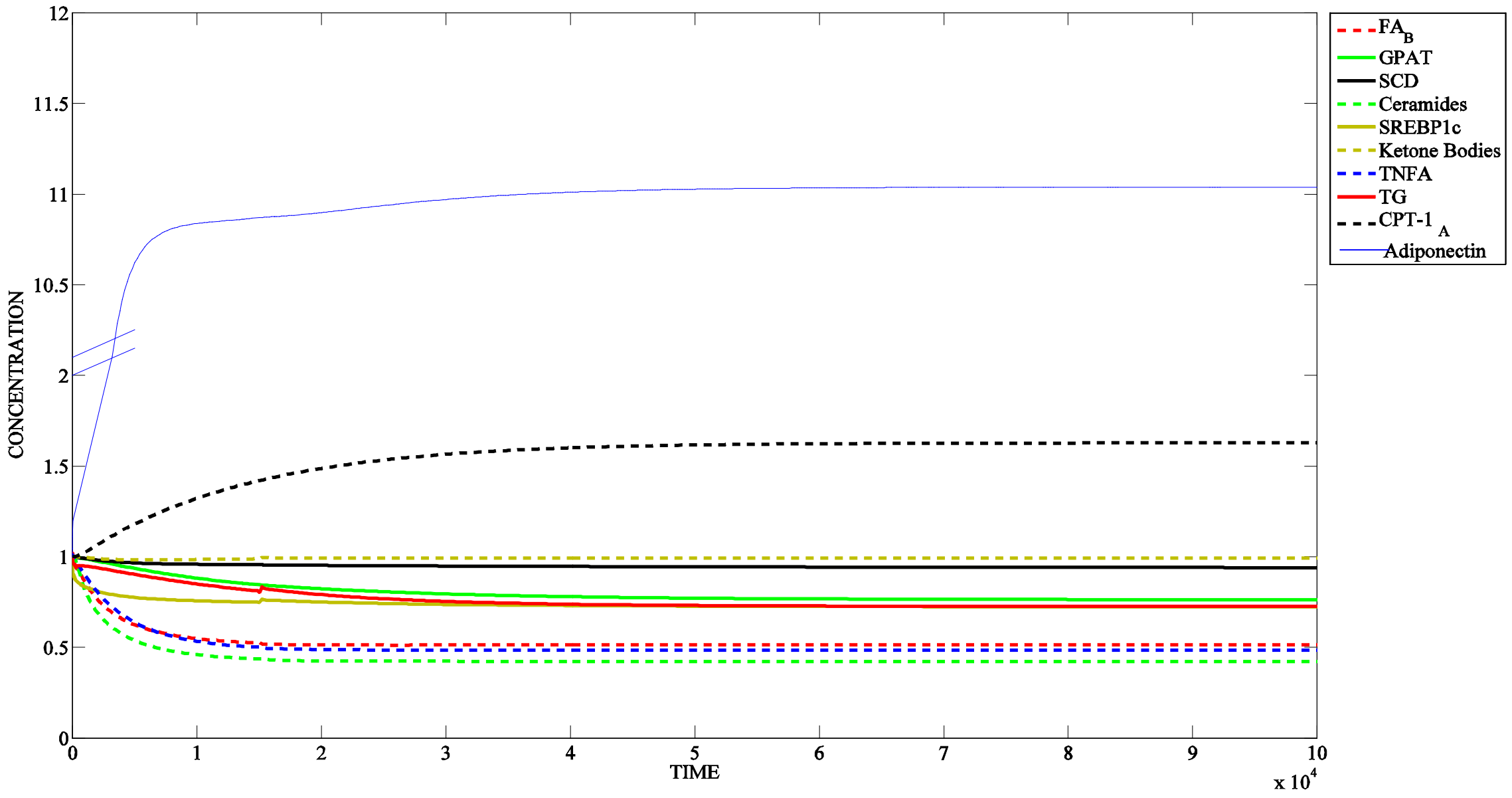

**Figure 5. Model validation of adiponectin overexpression under fasting conditions.** Overexpression of adiponectin under fasting conditions was simulated by decreasing the rate of adiponectin degradation by 10-fold, resulting in increased levels of adiponectin and the glucose influx was reduced by 10-fold. In concordance with experimental observations, increased adiponectin concentration results in downregulation of serum fatty acids ($FA_B$), hepatic triglycerides (TG), TNF-α, ceramides, SREBP1c, GPAT and SCD and increased expression of adipose CPT-1 ($CPT\text{-}1_A$), with no changes in serum ketone bodies concentration.

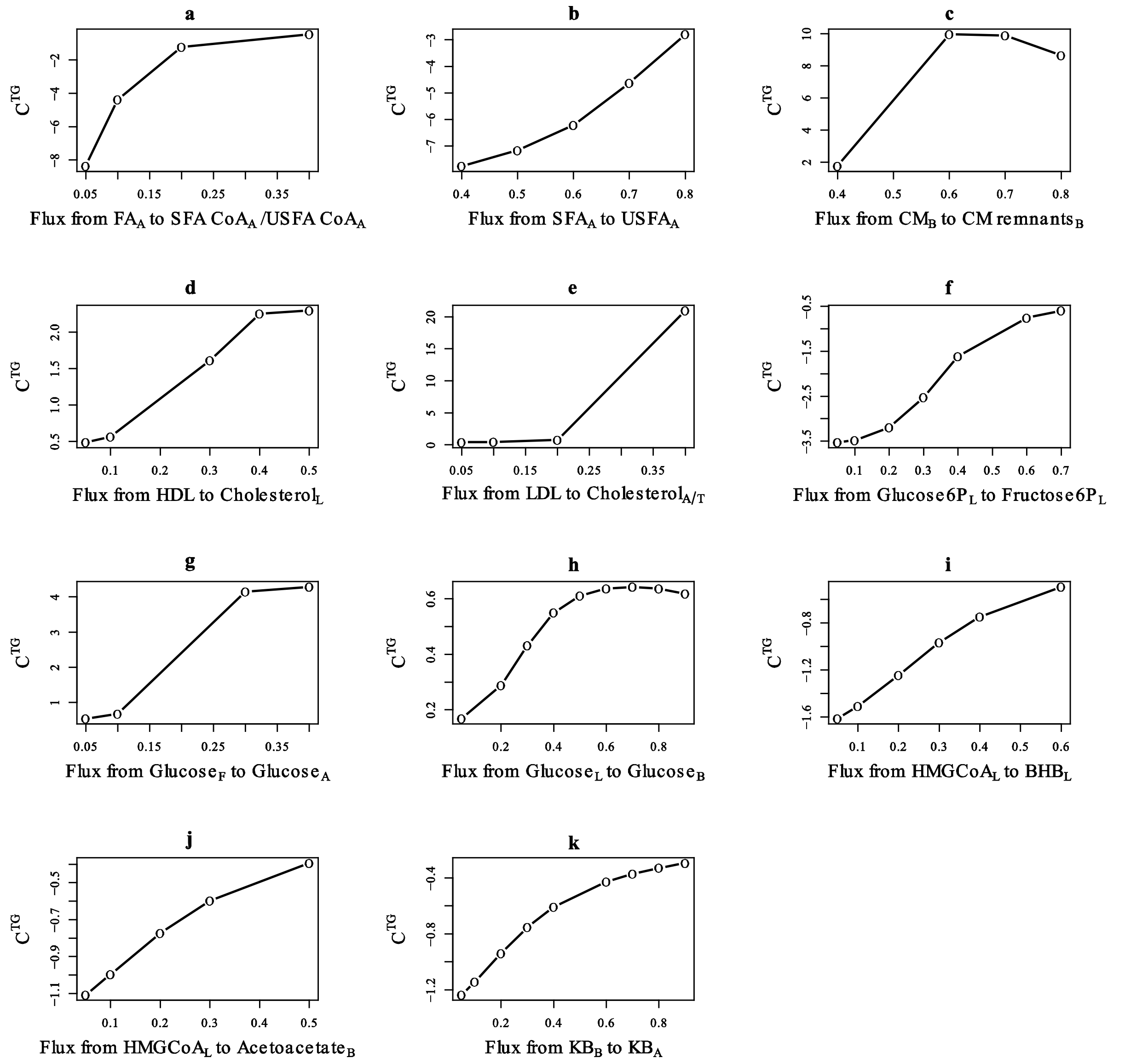

**Figure 6. High dynamic concentration control coefficients of branch-points SteatoNet.** Range of concentration control coefficients of a) activation of saturated (SFA) and unsaturated (USFA) fatty acids in adipose, b) desaturation of SFA into USFA in adipose, c) breakdown of chylomicron (CM) into chylomicron remnants, d) reverse cholesterol transport, e) Low density lipoprotein (LDL) distribution to adipose and peripheral tissues, f) fructose-6-phosphate synthesis from glucose-6-phosphate, g) glucose transport to adipose, h) hepatic release of glucose into blood, i) β-hydroxybutyrate (BHB) synthesis from 3-hydroxy 3-methylglutaryl coenzyme A (HMG CoA), j) acetoacetate transport to blood, and k) uptake of ketone bodies (KB) by adipose,; with respect to hepatic triglyceride concentration.

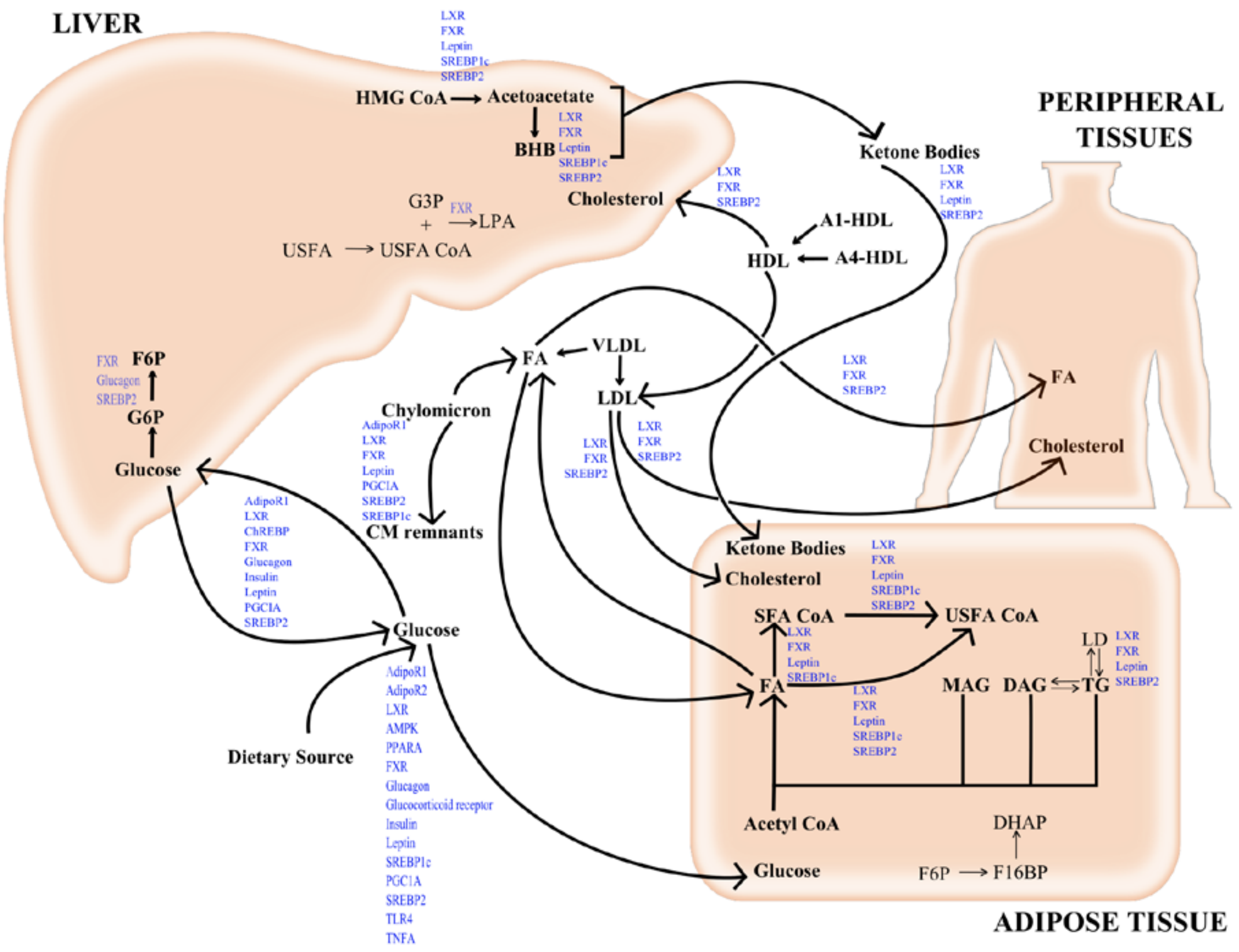

**Figure 7. Pathway branch-points and regulatory factors with high concentration control coefficients to hepatic triglyceride sensitivity.** Concentration control coefficients ($C_f^{TG}$) of pathway branch-points were calculated with respect to hepatic triglyceride concentration and the concentration of various transcriptional and post-translational pathway regulators, by altering the initial flux distribution parameter ($f$) of each branch-point in the network by intervals of 10% and determining the change in triglyceride or regulator concentration. The pathways in bold highlight high dynamic sensitivity branch-points while the others specify high-sensitivity+ low tolerance 'stiff' branch-points. Pathway regulators that are influenced by alterations in flux distribution are labelled at the corresponding branch-point to which they are sensitive. LXR- Liver X Receptor, SREBP-1c- Sterol Regulatory Element-Binding Protein-1c, TAG/TG- Triacylglycerol/ Triglycerides, ChREBP- Carbohydrate Response Element-Binding Protein, AMPK- Adenosine Monophosphate-regulated Protein Kinase,

VLDL- Very-Low Density Lipoprotein, LDL- Low density lipoprotein, SREBP-2- Sterol Regulatory Element-Binding Protein-2, TNFA- Tumour Necrosis Factor alpha, FA- Fatty Acids, G6P- Glucose-6-Phosphate, F6P- Fructose-6-Phosphate, USFA- Unsaturated Fatty Acids, G3P- Glycerol-3-Phosphate, LPA- Lysophosphatidic Acid, HMG CoA- 3-Hydroxy 3-Methylglutaryl Coenzyme A, BHB- Beta-Hydroxybutyrate, HDL- High-density Lipoprotein, CM- Chylomicron, SFA- Saturated Fatty Acids, MAG- Monoacylglycerol, DAG-Diacylglycerol, LD- Lipid Droplets, DHAP- Dihydroxyacetone Phosphate, F16BP- Fructose-1-6 Biaphosphate, Adipo R1/R2- Adiponectin Receptor 1/2, PPARA- Peroxisome Proliferator-Activated Receptor alpha, FXR- Farnesoid X Receptor, PGC1A- Peroxisome Proliferator-Activated Receptor gamma Coactivator 1 alpha, TLR4- Toll-like Receptors 4, PXR- Pregnane X Receptor.

**SUPPLEMENTARY INFORMATION**

**Supplementary Figure 1. Complete set of equations describing metabolic flux distribution at pathway branch points in the SteatoNet liver compartment.**

| **Pathway branch** | | |
|---|---|---|
| Blood_glucose -> Adipose_glucose | 0.4<br>0.2 | Hepatic_glucose -> Blood_glucose<br>Glucose_source |
| Blood_glucose -> Tissue_glucose | 0.4<br>0.2 | Hepatic_glucose -> Blood_glucose<br>Glucose_source |
| Hepatic_glucose -> Blood_glucose | 0.1<br>0.8 | Blood_glucose -> Hepatic_glucose<br>Glucose-6-P -> Hepatic_glucose |
| Glucose-6P -> Glucose-1P | 0.2 | Hepatic_glucose -> Glucose-6P |
| Glucose-6P -> Fructose-6P | 0.5 | Hepatic_glucose -> Glucose-6P |
| Glucose-6P -> Ribulose-5P | 0.05 | Hepatic_glucose -> Glucose-6P |
| Glycogen -> Glucose-1P | 0.5 | UDP-Glucose -> Glycogen |
| Fructose-6P -> Fructose-2,6BP | 0.1 | (Glucose-6P -> Fructose6P) + (Fructose-1,6BP -> Fructose-6P) + (Fructose-2,6BP -> Fructose-6P) + (Xylulose-5P + Erythrose-4P -> Glyceraldehyde-3P + Fructose-6P) + (Sedoheptulose-7P + Glyceraldehyde-3P -> Fructose-6P + Erythrose-4P) |
| Fructose-1,6BP -> DHAP | **0.2** | Fructose-6P -> Fructose-1,6BP |
| Fructose-1,6BP -> Glyceraldehyde-3P | **0.2** | Fructose-6P -> Fructose-1,6BP |
| DHAP -> Glycerol-3P | **0.1** | Fructose-1,6BP -> DHAP |
| Lactate -> Pyruvate | **0.3**<br>**0.05** | **Lactate_source**<br>Pyruvate -> Lactate |
| Pyruvate -> Oxaloacetate | **0.1** | (PEP -> Pyruvate) + (Lactate -> Pyruvate) + (Serine -> Pyruvate + Ammonia) + (Cysteine -> Pyruvate + Ammonia) |
| | | + (Malate -> Pyruvate) |
| Pyruvate -> mito_AcetylCoA | **0.1** | (PEP -> Pyruvate) + (Lactate -> Pyruvate) + (Serine -> Pyruvate + Ammonia) + (Cysteine -> Pyruvate + Ammonia) + (Malate -> Pyruvate) |
| Oxaloacetate -> PEP | **0.3** | (Pyruvate -> Oxaloacetate) + (Citrate -> AcetylCoA + Oxaloacetate) + (Malate -> Oxaloacetate) |
| AcetylCoA -> cyto_HMGCoA | **0.1** | (Citrate -> AcetylCoA + Oxaloacetate) + (MalonylCoA -> cyto_AcetylCoA) |
| MalonylCoA -> cyto_AcetylCoA | **0.1** | cyto_AcetylCoA -> MalonylCoA |
| Glycerate-3P -> Phospho-3-hydroxypyruvate | **0.7** | Glyceraldehyde -> -> -> Glycerate2P |
| Serine -> Glycine | **0.1** | (Phospho-3-serine -> Serine) + Serine_source |
| Serine -> Pyruvate + Ammonia | **0.5** | (Phospho-3-serine -> Serine) + Serine_source |
| Cysteine -> Pyruvate + Ammonia | **0.3** | (Cystathionine -> Cysteine) + Cysteine_source |
| Methionine -> Homocysteine | **0.4** | (Homocysteine -> Methionine) + Methionine_source |
| Homocysteine + Serine -> Cystathionine | **0.1** | (Phospho-3-serine -> Serine) + Serine_source |
| Phenylalanine -> Tyrosine | **0.1** | Phenylalanine_source |
| Tyrosine -> Acetoacetate + Fumarate | **0.1** | (Phenylalanine -> Tyrosine) + Tyrosine_source |
| Oxoglutarate + Ammonia -> Glutamate | **0.3**<br>**1.0**<br><br>**1.0**<br>**1.0**<br><br>**1.0** | Oxalosuccinate -> Oxoglutarate<br>Oxaloacetate + Glutamate -> Aspartate + Oxoglutarate<br>Pyruvate + Glutamate -> Alanine + Oxoglutarate<br>Glutamic_Semialdehyde + Glutamine -> Arginine+ Oxoglutarate<br>Glutamic_Semialdehyde + Glutamate -> Ornithine + Oxoglutarate |
| Glutamate -> -> Glutamic_Semialdehyde | **0.05** | (Oxoglutarate + Ammonia -> Glutamate) + Glutamate_source + (Aspartate + Glutamine -> Asparagine + Glutamate) + (Glutamine -> Glutamate + Ammonia) + |
| | | (Histidine -> Glutamate) |
| Glutamate + Ammonia -> Glutamine | **0.5** | (Oxoglutarate + Ammonia -> Glutamate) + Glutamate_source + (Aspartate + Glutamine -> Asparagine + Glutamate) + (Glutamine -> Glutamate + Ammonia) + (Histidine -> Glutamate) |
| Glutamine -> Glutamate + Ammonia | **0.7** | Glutamine_source + (Glutamate + Ammonia -> Glutamine) |
| Pyruvate + Glutamate -> Alanine + Oxoglutarate | **0.3** | (PEP -> Pyruvate) + (Lactate -> Pyruvate) + (Serine -> Pyruvate + Ammonia) + (Cysteine -> Pyruvate + Ammonia) + (Malate -> Pyruvate) |
| Aspartate + Glutamine -> Asparagine + Glutamate | **0.1** | (Oxaloacetate + Glutamate -> Aspartate + Oxoglutarate) + Aspartate_source |
| Glutamic_Semialdehyde + Glutamine -> Arginine + Oxoglutarate | **0.1** | Glutamate ->-> Glutamic_Semialdehyde |
| Isoleucine -> PropionylCoA | **0.1** | Isoleucine_source |
| Valine_to_PropionylCoA | **0.1** | Valine_source |
| Alpha_Ketobutyrate -> PropionylCoA | **0.1** | Cystathionine -> Cysteine |
| PropionylCoA -> MethylmalonylCoA | **0.1** | (Isoleucine -> PropionylCoA) + (Valine_to_PropionylCoA) + (Alpha_Ketobutyrate -> PropionylCoA) |
| Histidine -> Glutamate | **0.1** | Histidine_source |
| Leucine -> HMGCoA | **0.1** | Leucine_source |
| Tryptophan -> Alanine + Hydroxy_3_Anthranilate | **0.1** | Tryptophan_source |
| Oxaloacetate + Glutamate -> Aspartate + Oxoglutarate | **0.4** | (Pyruvate -> Oxaloacetate) + (Citrate -> AcetylCoA + Oxaloacetate) + (Malate -> Oxaloacetate) |
| Arginine -> Urea + Ornithine | **0.9** | Arginine_source + (Glutamic_Semialdehyde + Glutamine -> Arginine + Oxoglutarate) |
| Glutamic_Semialdehyde + | **0.1** | Glutamate ->-> Glutamic_Semialdehyde |
| Glutamate -> Ornithine + Oxoglutarate | | |
| Carbamoyl_phosphate + Ornithine -> Citrulline | **0.2** | (Glutamic_Semialdehyde + Glutamate -> Ornithine + Oxoglutarate) + (Arginine -> Urea + Ornithine) |
| Citrate -> CisAconitate | **0.6** | mito_AcetylCoA + Oxaloacetate -> Citrate |
| Malate -> Oxaloacetate | **0.8**<br>**0.1** | Fumarate -> Malate<br>Oxaloacetate -> Malate |
| Keto-3-AcylCoA -> AcylCoA | **0.2** | Trans2BEnoylCoA -> Keto-3-AcylCoA |
| SuccinylCoA + Acetoacetate -> Succinate + AcetoacetylCoA | **0.1** | SSuccinylDihydrolipamideE1 -> SuccinylCoA |
| mito_AcetylCoA + Oxaloacetate -> Citrate | **0.2** | (Pyruvate -> Oxaloacetate) + (Citrate -> AcetylCoA + Oxaloacetate) + (Malate -> Oxaloacetate) |
| Acetoacetate -> blood_Acetoacetate | **0.4**<br>**0.2** | HMGCoA -> Acetoacetate<br>Tyrosine -> Acetoacetate + Fumarate |
| Acetoacetate -> BHydroxybutyrate | **0.5**<br>**0.2** | HMGCoA -> Acetoacetate<br>Tyrosine -> Acetoacetate + Fumarate |
| blood_BHydroxybutyrate -> adipo_ BHydroxybutyrate | **0.5** | BHydroxybutyrate -> blood_ BHydroxybutyrate |
| blood_Acetoacetate -> adipo_ Acetoacetate | **0.5** | Acetoacetate -> blood_ Acetoacetate |
| Palmitate -> Keto-3-Sphingosine | **0.01** | (MalonylCoA + AcetylCoA -> Palmitate) + (blood_Palmitate -> Palmitate) |
| PalmitoylCoA -> PalmitoleateCoA | **0.9** | Palmitate -> PalmitoylCoA |
| PalmitoleoylCoA + Glycerol3P -> LPA | **0.05** | (Palmitoleate -> PalmitoleoylCoA) + (PalmitoylCoA -> PalmitoleoylCoA) |
| DAG + Choline -> PC | **0.2** | (TG -> DAG) + (PA -> DAG) + (MAG -> DAG) |
| DAG -> MAG | **0.1** | (TG -> DAG) + (PA -> DAG) + (MAG -> DAG) |
| PalmitoleateCoA + DAG -> TG | **0.5** | (TG -> DAG) + (PA -> DAG) + (MAG -> DAG) |
| TG_to_DAG | **0.3** | (PalmitoleateCoA + DAG -> TG) + (blood_TG -> TG) + |
| | | (TG_lipid_droplet -> TG) |
| TG -> VLDL | **0.7** | (Cholesterol + PalmitoleoylCoA -> Cholesteryl_esters) + (Cholesterol + PalmitoylCoA -> Cholesteryl_esters) |
| TG -> TG_lipid_droplet | **0.3** | (PalmitoleoylCoA + DAG -> TG) + (blood_TG -> TG) + (TG_lipid_droplet -> TG) |
| MAG -> Glycerol | **0.8** | DAG -> MAG |
| PC -> PA + Choline | **0.2** | DAG + Choline -> PC |
| Cholesterol + PalmitoylCoA -> Cholesteryl_esters | **0.05** | Palmitate -> PalmitoylCoA |
| Cholesterol + PalmitoleoylCoA -> Cholesteryl_esters | **0.1** | (Palmitoleate -> PalmitoleoylCoA) + (PalmitoylCoA -> PalmitoleoylCoA) |
| Cholesterol_utilization | **0.7** | Cholesterol_source + (LDL_cholesterol -> Cholesterol) + (A_2_HDL -> Cholesterol) + (HMGCoA -> Cholesterol) |
| blood_Cholesterol -> macrophage_Cholesterol | **0.3** | (VLDL -> blood_fatty_acids) + (HDL -> VLDL + LDL) |
| blood_Cholesterol -> adipo_Cholesterol | **0.3** | (VLDL -> blood_fatty_acids) + (HDL -> VLDL + LDL) |
| blood_Cholesterol -> tissue_Cholesterol | **0.3** | (VLDL -> blood_fatty_acids) + (HDL -> VLDL + LDL) |
| A_2_HDL -> Cholesterol | **0.2** | (A_4_HDL -> A_2_HDL + A_3_HDL) + (A_1_HDL -> A_2_HDL + VLDL) |
| blood_Fatty_acids -> Palmitoleate | **0.1** | (adipo_Fatty_acids -> blood_Fatty_acids) + (VLDL -> blood_fatty_acids) + (Chylomicron_blood -> blood_Fatty_acids) |
| Chylomicron -> Chylomicron_remnants | **0.5** | Chylomicron_source |